# Low loss spin wave resonances in organic-based ferrimagnet vanadium tetracyanoethylene thin films

Na Zhu,[1] Xufeng Zhang,[1] I.H. Froning,[2] Michael E. Flatté,[3] E. Johnston-Halperin,[2] and Hong X. Tang[1]

[1]*Department of Electrical Engineering, Yale University, New Haven, Connecticut 06511, USA*

[2]*Department of Physics, The Ohio State University, Columbus, Ohio 43210-1117, USA*

[3]*Optical Science and Technology Center and Department of Physics and Astronomy, University of Iowa, Iowa City, Iowa 52242, USA*

Abstract:

We experimentally demonstrate high quality factor spin wave resonances in an encapsulated thin film of the organic-based ferrimagnet vanadium tetracyanoethylene ($V[TCNE]_{x\sim2}$) coated on an a-plane sapphire substrate by low temperature chemical vapor deposition. The thickness standing wave modes are observed in a broad frequency range (1 GHz ~ 5 GHz) with high quality factor exceeding 3,200 in ambient air at room temperature, rivaling those of inorganic magnetic materials. The exchange constant of $V[TCNE]_{x\sim2}$, a crucial material parameter for future study and device design of the $V[TCNE]_{x\sim2}$, is extracted from the measurement with a value of $(4.61 \pm 0.35) \times 10^{-16}$ m$^2$. Our result establishes the feasibility of using organic-based materials for building hybrid magnonic devices and circuits.

Spin waves, which are the collective excitation of the magnetization in magnetic materials, have been attracting intensive attention recently due to potential applications in both fundamental research[1–4] and device applications[5–9] thanks to properties such as being ohmic loss free,[10,11] long spin lifetime, and large-bandwidth tunability. In particular, yttrium iron garnet (YIG, $Y_3Fe_5O_{12}$) has been long considered as one of the most attractive magnon media, thanks to its magnetic, microwave, mechanical and optical properties. As a result, YIG has been widely adopted to investigate the interactions among spin waves, microwaves[12–19], acoustic waves[20,21] and optical excitations.[22–26] However, high quality YIG films can only been grown on specific lattice-matched substrates such as gadolinium gallium garnet (GGG,

$Ga_3Gd_5O_{12}$),[27] so its integration with other substrates, such as silicon, is extremely difficult. Moreover, micro/nano patterning has long been an obstacle in YIG fabrication, which significantly limits the development and applications of YIG-based magnonic circuits.

To overcome these constraints, organic-based magnets are emerging as an alternative solution. Such materials can be grown as high quality thin films on various substrates,[28–30] have excellent post fabrication capabilities for developing complex geometries, and require only low temperature and ambient pressure deposition conditions.[31] Particularly, vanadium tetracyanoethylene (V[TCNE]$_{x\sim2}$),[32] an organic-based ferrimagnetic semiconductor ($E_g = 0.5$ eV, $\sigma = 0.01$ S/cm),[28] exhibits room temperature magnetic ordering ($T_C > 600$ K) with a narrow ferromagnetic resonance linewidth (peak to peak linewidth of 1G),[31] which is comparable to that of YIG.[33] These advantages suggest that V[TCNE]$_{x\sim2}$ could be a substitute for YIG in developing high-quality magnonic circuits, especially for integration on a chip. However, the saturation magnetization of V[TCNE]$_{x\sim2}$ (95 G) is over an order of magnitude smaller compared with that of YIG (1850 G), which will offer potential applications in low-current spin-transfer switching,[34] while also introduce the limitations in the practical devices design in diverse areas such as high-density magnetic recording, power transformer and electromagnetic interface (EMI) prevention components.[35,36]

Although magnetic excitations can take place in many different and complex forms, previous studies on V[TCNE]$_{x\sim2}$ have mainly focused on the characterization of its ferromagnetic resonance (FMR). The studies of magnetostatic waves and exchange interactions, which are very common in YIG thin film devices, have thus far been overlooked. Magnetostatic waves, in contrast to the FMR mode in which all the spins precess uniformly, are propagating spin waves that have non-zero wavevectors and non-uniform field distributions in the material.[37]

Unlike the FMR mode, which resonates at a single frequency, magnetostatic waves can exist in a frequency range depending on their mode profiles. In particular, it has been demonstrated that in ferromagnetic thin films magnetostatic waves can form a series of standing wave modes due to spin pinning at the film surfaces.[38,39] These spin wave resonances have potential applications such as serving as high quality factor magnonic multi-mode resonators for information storage and processing.

In this paper, spin wave resonances along the film thickness direction in a V[TCNE]$_{x\sim2}$ thin film are investigated. Standing waves with different mode orders in both perpendicularly and in-plane magnetized thin films are recorded in a broad frequency range from 1 GHz to 5 GHz that show very high quality ($Q$) factors, exceeding $Q = 3200$, indicating high material quality and thickness uniformity of this organic-based ferrimagnetic thin film. The exchange constant of V[TCNE]$_{x\sim2}$ can be extracted from the free spectral range (FSR) of the standing wave modes, which is a crucial material parameter for the potential device applications of V[TCNE]$_{x\sim2}$. Our findings advance the study on such magnetic polymers from the previous simple FMR characterization, and establish the feasibility of their potential use for developing functional spin wave devices.

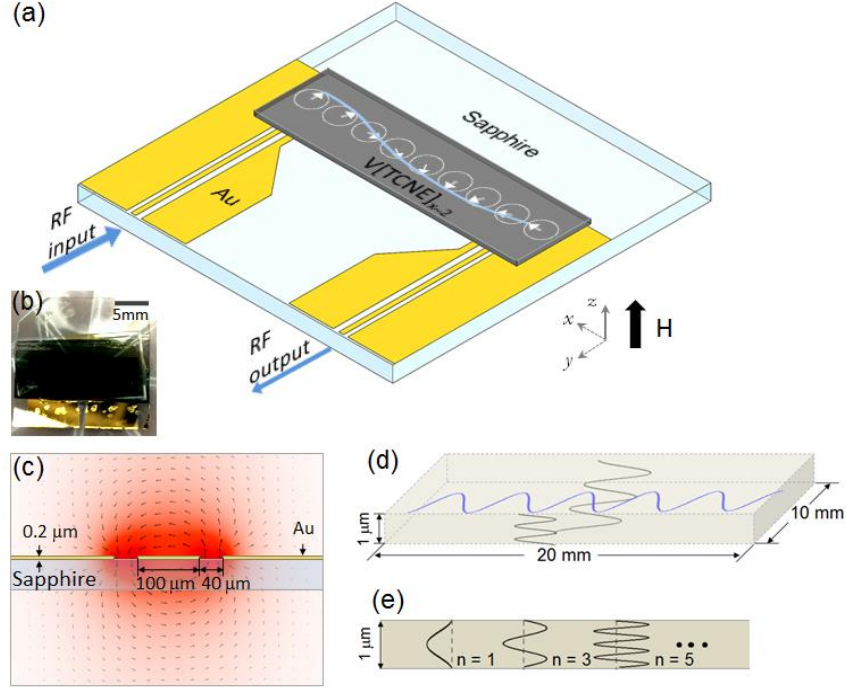

FIG. 1. (a) Schematic of the V[TCNE]$_{x\sim2}$ magnonic waveguide used in this experiment. (b) An encapsulated V[TCNE]$_{x\sim2}$ thin film on a sapphire substrate patterned with gold transmission lines. (c) Simulated magnetic field distribution of the coplanar waveguide. The black arrows and colors indicate the magnetic field directions and amplitudes, respectively. (d) Schematic illustration of spin wave resonances along length, width and thickness directions. (e) Standing wave modes with odd mode number in a 1-μm-thick V[TCNE]$_{x\sim2}$ thin film.

The V[TCNE]$_{x\sim2}$ device used in our experiments is illustrated in Fig. 1(a), and consists of a pair of gold coplanar waveguides and a 1-μm-thick V[TCNE]$_{x\sim2}$ film. The device is wire-bonded to a FR-4 printed circuit board for microwave measurement, which is not shown. The two 15-mm-long coplanar transmission lines are designed with a 100-μm-wide center signal line and a 40-μm-wide gap between the signal line and the ground, and a characteristic impedance of 50 Ω. These two coplanar waveguides serve as the transducer for spin wave excitation and detection, and are separated by 5 mm from each other. To fabricate these coplanar waveguides, photolithography is first carried out on a 20×20×0.43 mm$^3$ a-plane (100) sapphire substrate using bi-layer resist process (S1813 photoresist on top of LOR 5A resist). After exposing in EVG 620 Mask Aligner, the sample is developed in MF319. The prepatterned sample is then deposited with 200-nm-thick gold, and a following lift-off process eventually transfers the pattern to the gold layer and forms the transducer chip, as shown in Fig. 1(b). The microwave magnetic field distribution of the designed coplanar waveguide is shown in Fig. 1(c).

The V[TCNE]$_{x\sim2}$ film is directly grown on the fabricated coplanar circuit to a thickness of 1 μm via low temperature (50°C) chemical vapor deposition in an argon atmosphere. It is deposited into a 20×10 mm$^2$ mesa bridging the two coplanar transmission lines using a shadow mask, and serves as the magnonic waveguide. TCNE and V(CO)$_6$ are used in the ratio of 10:1 at evaporation temperatures of 50°C and 10°C, respectively, yielding a deposition rate of 9.3 nm/min.[40] Because of its air-sensitivity, the film is encapsulated using a technique that preserves the magnetic order for over one month under ambient condition.[41] The technique requires applying an epoxy (purchased from Ossila) to the film surface. A glass cover slide is placed over the epoxy and pressed down firmly to spread the epoxy across the film surface and edges to isolate the film from the air, and to provide mechanical protection for the sample. In a final step, the epoxy is cured by a white LED for over an hour.

The fabricated device is characterized by the microwave transmission measurement after wire-bonding to a FR-4 printed circuit board terminated with SMA connectors for microwave input and output. The input signal is sent to the device from one port of the vector network analyzer (VNA), and the output signal is amplified by a radio frequency (RF) amplifier to compensate the insertion loss of the device, which we attribute to the low transducer efficiency associated with the relatively small spin density of the material. When a bias magnetic field is applied to the device, magnetostatic waves can be excited in a given frequency range and propagate in the V[TCNE]$_{x\sim2}$ magnonic waveguide. Ideally, such transmission windows would show up in the transmission spectra as a series of peaks, which correspond to the discrete standing wave modes formed due to the confinement of the film boundaries. However, in our experiment, the electrical cross talk between the two transmission lines is significant due to the low spin density of the magnonic media, and gives rise to a transmission background which distorts the resonances into a Fano line shape, sometimes even Lorentzian dips, instead of the ideal Lorentzian peak line shape.

Due to the finite size of ferromagnetic thin films, it is possible that the spins near the film surfaces are not able to precess due to the different anisotropy fields at the surface compared with those in the bulk. In this condition, the reflection of propagating spin waves at surfaces can form a series of standing wave modes along the thickness, length and width directions with different FSRs, as shown in Fig. 1(d). The magnetostatic resonances formed by volume waves with non-zero in-plane wavevectors have been extensively studied and observed in YIG thin films and permalloy metallic thin films.[42,43] For the volume waves in a tangentially magnetized thin film when the direction of magnetic field in along length direction (axis $x$ in Fig. 1(a)), the dispersion relation is described by[44]

$$(1 + \eta^2) + 2|(1+\eta^2)^{1/2}|\left(-\frac{1+\eta^2+\kappa}{1+\kappa}\right)^{1/2} \times (1+\kappa)\cot\left[|k_y|d\left(-\frac{1+\eta^2+\kappa}{1+\kappa}\right)^{1/2}\right]$$

$$+(1+\kappa)^2\left(-\frac{1+\eta^2+\kappa}{1+\kappa}\right) - \nu^2 = 0, \quad (1)$$

where $\eta = k_x/k_y$, $\Omega_H = H_0/4\pi M_s$, $\Omega = \omega/4\pi\gamma M_s$, $\kappa = \Omega_H/(\Omega_H^2 - \Omega^2)$, and $\nu = \Omega/(\Omega_H^2 - \Omega^2)$. Here $k_x$ and $k_y$ are the wavevectors along length and width directions, $d$ is film thickness, $H_0$ is the effective magnetic field inside the film, and $4\pi M_s$ is the saturation magnetization, which is 95 G according to previous characterization of the V[TCNE]$_{x\sim2}$.[31] $\gamma$ is gyromagnetic ratio with the value of 2.8 MHz/Oe from the previous study,[31] and $\omega$ is the resonance frequency. Based on the dimension of the deposited V[TCNE]$_{x\sim2}$ thin film, the calculated FSRs along the length and width directions are around 20 kHz and 50 kHz, respectively. As a result, the longitudinal and lateral standing wave modes are packed very close together with FSRs much smaller than the typically ultra-narrow ferromagnetic resonance linewidth (~ 2 MHz), and therefore are unresolvable in the spectrum. Similarly, for the perpendicularly magnetized thin film, the frequency spacings of in-plane standing wave modes can be calculated from the dispersion relation,[37] and the theoretical FSRs are 12.5 kHz and 25.1 kHz for the modes along length and width directions, respectively, which are also too small to be resolved.

Based on the discussion above, under the surface pinning condition, only spin wave resonances along the thickness direction are resovable due to the small film thickness, $d$. The resonances occur whenever the film thickness equals to an integral number of half wavelengths, but only odd modes can be excited by a uniform RF field,[38] as shown in Fig. 1(e). The resonant frequencies are determined by

$$f_n = f_0 + f_m \lambda_{\text{ex}} \left(\frac{n\pi}{d}\right)^2, \tag{2}$$

where $f_0$ equals $\gamma(H_0 - 4\pi M_s)$ and $\gamma\sqrt{H_0(H_0 + 4\pi M_s)}$ for normally and tangentially magnetized thin film, respectively.[37] Here $H_0$ is the applied bias magnetic field, $4\pi M_s$ is the saturation magnetization, $d$ is film thickness, $\lambda_{\text{ex}}$ is the exchange constant, and $f_m = 4\pi\gamma M_s$.

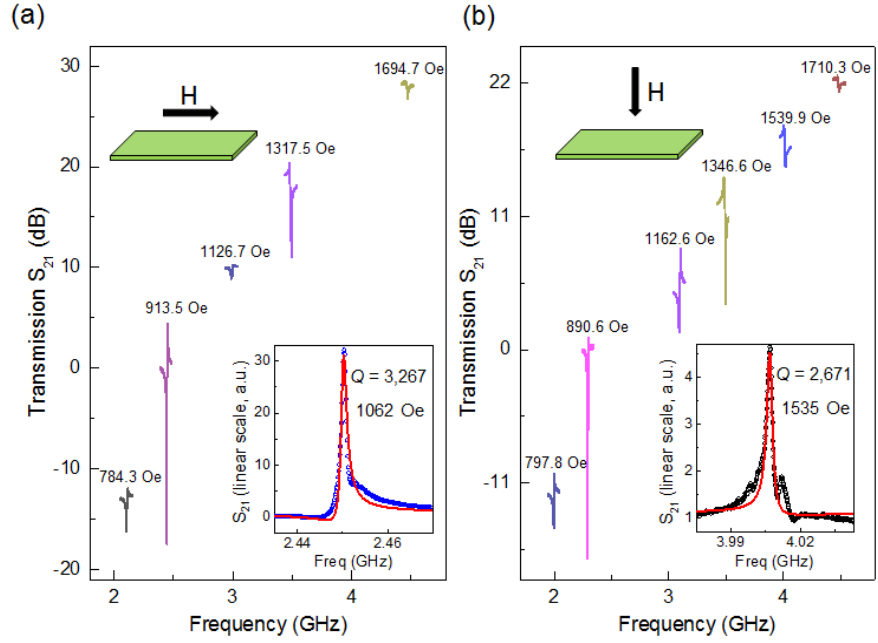

FIG. 2. Vector network analyzer transmission characterization of the V[TCNE]$_{x\sim2}$ magnonic waveguide with magnitude response at different bias magnetic fields when the film is tangentially (a) and perpendicularly (b) magnetized. Inset: zoom-in microwave transmission spectra (symbols) and Fano line shape fitting (lines) of spin wave resonances under the uniform bias magnetic field.

Figures 2 (a) and (b) show the normalized microwave transmission spectra when the film is tangentially and normally magnetized, respectively. By varying the magnitude of the bias magnetic field, spin wave resonances with Fano line shape are observed with frequencies determined by the magnitude of the applied magnetic field, which are attributed to the standing wave modes along thickness direction, as discussed above. Clear fundamental ($n = 1$) modes are observed in a large frequency range (1~5 GHz). The additional features near the resonances with smaller extinction ratio are attributed to high order modes which have relatively lower excitation and detection efficiencies.

The spin wave resonances have very narrow linewidth and large amplitude extinction ratios, which indicate the high quality of the V[TCNE]$_{x\sim2}$ thin film. The $Q$ factor of the transmission peaks is defined as $f/\delta f$, where $f$ is the center frequency and $\delta f$ is the full width at half maximum (FWHM). As shown in the inset of Fig. 2(a), Fano line shape fitting is applied for the resonance with center frequency at 2.45 GHz when the thin film is tangentially magnetized. $Q$ factors as high as 3,267 are observed, corresponding to a linewidth $\delta f = 0.75$ MHz, comparable to that of FMR mode in YIG thin films.[27,33] Similarly, the high $Q$ factor ($Q = 2671$) is also achieved in the normal magnetization condition for the first order spin wave resonance centered at 4 GHz, as shown in the inset of Fig. 2(b).

These high Q resonances which are comparable with those of YIG can be explained considering the typical damping mechanisms of spin-waves. In both YIG and V[TCNE]$_{x\sim2}$, damping is caused by magnetic dipole coupling and therefore it is proportional to the saturation magnetization; for example, surface imperfections permit decay of a spin wave into two other spin waves (a magnetic Raman process).[45,46] As the saturation magnetization of VTCNE is over an order of magnitude smaller than YIG, the limiting damping from these processes could be significantly smaller than in YIG.

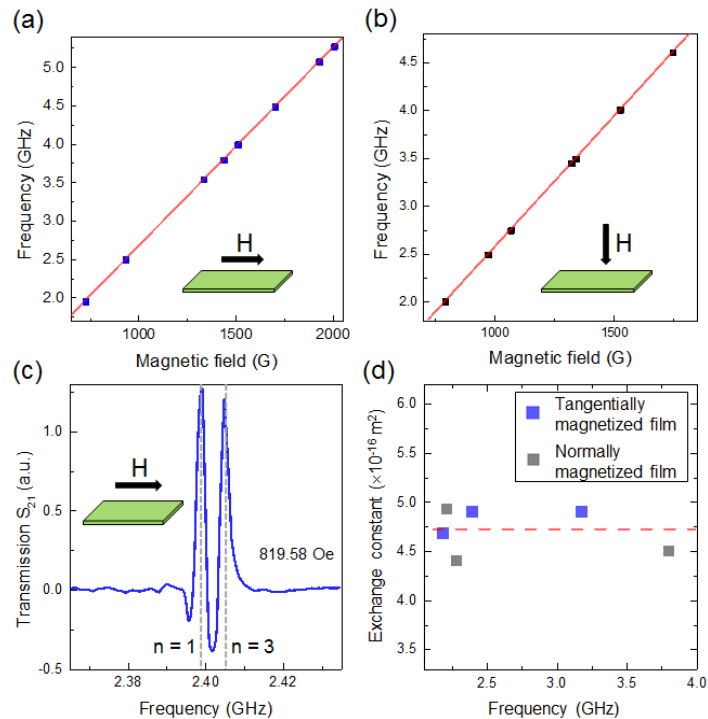

FIG. 3. The f–H relation for both tangentially (a) and perpendicularly (b) magnetized thin films is fitted by Eq. (2) to extract gryomagnetic ratio and saturation magnetization. (c) Transmission spectrum of the in-plane magnetized thin film shows two resonances with different mode numbers. (d) Extracted exchange constant of the V[TCNE]$_{x\sim2}$ thin film at different frequencies, in both normal and tangential magnetization configurations. The red dash line denotes the mean value of the extracted exchange constant.

The f-H relations can be extracted for both tangentially and normally magnetized thin films using Eq. (2), and are plotted in Figs. 3(a) and (b). The gyromagnetic ratio $\gamma$ is fitted with the values of 2.59 MHz/Oe and 2.73 MHz/Oe for tangential and normal magnetization configurations, respectively. The fitted values in our experiments are close to the value reported previously (2.8 MHz/Oe). The fitted values for the saturation magnetization for tangentially and normally magnetized thin films are 56.15 G and 60.32 G, respectively. These fitted values of the saturation magnetization are slightly smaller compared with the previous studies of V[TCNE]$_{x\sim2}$ thin films which demonstrated 95 G saturation magnetization, perhaps resulting from the degradation of the spin density of this organic-based ferrimagnet due to the ambient measurement condition.[31]

Fig. 3(c) shows the spectrum of spin wave resonances at 819.58 Oe, where two clear resonances peaks with 5.625 MHz frequency spacing can be clearly resolved in this zoom-in spectrum. As we discussed above, only odd modes can be excited by a uniform RF field. Therefore, these two modes in the spectrum are attributed to the first and third order standing wave modes along thickness direction. The number of modes we can observe is mainly limited by the narrow excitation bandwidth of the coplanar waveguide. These two modes are observed in both normal and tangential magnetization configurations and exist at a wide frequency range with frequency spacing around 5.625 MHz. Based on Eq. (2), the exchange constant of the V[TCNE]$_{x\sim2}$ thin film can be extracted from the frequency spacing between these two modes. The values of exchange constant extracted at different frequencies are plotted in Fig. 3(d). The average value of calculated exchange constant is $(4.61 \pm 0.35) \times 10^{-16}$ m$^2$. This value is comparable to that of YIG ($3 \times 10^{-16}$ m$^2$),[37] demonstrating the wide variety of potential applications of the V[TCNE]$_{x\sim2}$ materials to be used as a substitute for YIG.

In summary, this work experimentally demonstrates the spin wave resonances in an organic-based ferrimagnet, V[TCNE]$_{x\sim2}$. The standing wave modes along the thickness direction are observed with narrow linewidth and $Q$ factors more than 3,200, which rival those of YIG despite the lack of crystalline structure in V[TCNE]$_{x\sim2}$, and meanwhile offer more flexibility in circuit design and fabrication thanks to the benefits of the benign deposition conditions of organic materials. Due to the low spin density of the organic films, the free spectral ranges of in-plane confined modes are too small to be resolved in the current device configuration. Further material and device development in patterning micro-scale waveguides may lead to the observation of in-plane magnetostatic wave modes. In our device fabrication processes, the direct sample growth on gold patterned sapphire substrates validates the outstanding properties of V[TCNE]$_{x\sim2}$ materials such as low deposition temperature, high rate, conformal coating on a wide variety of substrates and prepatterned circuits can be realized for ambient operation devices. Moreover, the exchange constant of V[TCNE]$_{x\sim2}$ materials is obtained from the frequency spacings between different orders of modes over a wide frequency range, which is a critical material parameter for future device and material studies of organic-based magnonic circuits. Our work presents the intriguing potential of this organic-based magnetic material to be used in making high $Q$ magnonic resonators for hybrid high frequency electronic and spintronic circuits.

This work is supported by DARPA/MTO MESO program and NSF Grant No. DMR-1507775. The authors acknowledge the NanoSystems Laboratory at Ohio State University. We would like to thank Michael Power and Christopher Tillinghast for the assistance in device fabrication.